\documentclass[final,5p,times,twocolumn]{elsarticle}
\usepackage[dvipsnames]{xcolor}
\usepackage[colorlinks=true]{hyperref}
\usepackage{siunitx}
\usepackage[figurename=Fig.,labelfont=bf,labelsep=period]{caption}
\usepackage{fancyhdr}
\usepackage{geometry}
\usepackage{bm}
\usepackage{amsmath,amsfonts,amsthm,amssymb}
\providecommand{\dod}[3][]{\mathinner{\dfrac{\mathrm{d}^{#1}#2}{\mathrm{d}^{\vphantom{#1}}{{#3}^{#1}}}}}

\providecommand{\dpd}[3][]{\mathinner{\dfrac{\partial^{#1}#2}{\partial^{\vphantom{#1}}{{#3}^{#1}}}}}
\newcommand{\lapl}[1][]{\vec{\nabla}^2_{\mkern-4mu{#1}\mkern3mu}}
\providecommand{\intj}[4][]{\int_{#2}^{#1} \! {#3} \, \mathrm{d} {#4}}
\newcommand{\abs}[1]{\lvert {#1} \rvert}
\graphicspath{{./figures/}}
\journal{}
\begin{document}

\pagestyle{fancy}
\fancyhf{}
\renewcommand{\headrulewidth}{0pt}
\fancyhead{}
\fancyhead[R]{\footnotesize{\thepage}}
\fancyhead[C]{\footnotesize{\textit{J. Zhang and Y. Zhang}}}
\fancyfoot{}

\begin{frontmatter}

\title{Local mobility variations along recrystallization boundaries}%

\author[nwu]{Jin Zhang}
\ead{jzhang@northwestern.edu}
\author[dtu]{Yubin Zhang\corref{cor1}}
\ead{yubz@dtu.dk}

\cortext[cor1]{Corresponding author}
\affiliation[nwu]{
  organization={Department of Materials Science and Engineering},
  addressline={Northwestern University}, 
  city={Evanston},
  postcode={60208}, 
  state={IL},
  country={USA}}
\affiliation[dtu]{
  organization={Department of Civil and Mechanical Engineering},
  addressline={Technical University of Denmark},
  city={Kgs. Lyngby},
  postcode={2800},
  country={Denmark}
}

\begin{abstract}
Understanding recrystallization boundary migration mechanisms is crucial for materials design. However, the lack of comprehensive mobility data for high-angle grain boundaries in typical polycrystalline samples has impeded gaining insights into the factors that govern boundary migration. Here, we propose an Iterative Mobility Matching (IMM) methodology to determine local mobility values and their spatial variations along recrystallization boundaries in partially recrystallized samples. This method optimizes mobility inputs in a phase field model by matching simulated microstructures with time-series recrystallization experiments. We first validate this method using synthetic data before applying it to real experimental data. For the first time, we reveal a three-order-magnitude variation in local mobility along individual recrystallization boundaries. The results show that local mobility depends strongly on boundary misorientation, and its variation significantly influences boundary dynamics, shedding light on the interplay of factors that impact the observed boundary migration behavior. The IMM methodology offers a new framework for interpreting experimental recrystallization behavior and can be broadly applied to other materials using \textit{in situ} experiments.
\end{abstract}

\begin{keyword}
  Recrystallization \sep Phase field \sep Grain boundary \sep Grain boundary mobility
\end{keyword}

\end{frontmatter}

\section{Introduction}\label{sec:intro}
Recrystallization is vital for optimizing material microstructure and enhancing properties \cite{Doherty1997}. This process involves the formation and growth of nearly defect-free nuclei, facilitated by the migration of recrystallization boundaries. It is crucial to understand how these boundaries migrate to control the microstructure and tailor the resulting material's performance. Unlike grain growth, where boundary migration is primarily driven by curvature, recrystallization involves boundary migration influenced by both curvature and stored energy. The local migration velocity of these boundaries, $v$, is governed by the equation:
\begin{equation}\label{eq:sharp}
  v = M F = M (F_{\text{s}} + F_\sigma),
\end{equation}
where $M$ is the mobility, and $F$ is the driving force for migration \cite{Haessner1978}. The driving force consists of the stored energy in the deformed matrix $F_{\text{s}}$ and the boundary-curvature-based driving force $F_\sigma=\sigma \kappa$. Here, $\sigma$ is the grain boundary energy, and $\kappa$ is the boundary curvature. The curvature driving force was overlooked in the literature but has recently been emphasized to be significant for accurately quantifying local boundary migration, as it can locally be as large as the stored energy \cite{Zhang2011,Zhang2014a,Moelans2013,Yadav2021}. 

It is critical to understand the role of different contributions to the boundary migration velocity to engineer the recrystallization microstructure. The local stored energy in the deformed matrix $F_{\text{s}}$ can be quantified by electron microscopes, and the boundary curvature $\kappa$ at discrete time steps can be determined from the experimental microstructure. However, despite its importance, there is limited research on the recrystallization boundary mobility. The mobility $M$ depends on the crystallography of the boundary and various physical factors. A comprehensive understanding of $M$ requires determining a large amount of mobility data. Current research primarily focuses on obtaining mobility values from boundary migration in bicrystals \cite{Gottstein2001,Molodov1994,Molodov2011,Winning2002} and using molecular dynamics \cite{Zhang2005,Holm2010,Olmsted2009,Homer2015}, which have mainly explored a limited number of boundaries with special misorientations and geometries. However, since the migration of these boundaries is not driven by stored energy, their applicability to recrystallization boundaries with arbitrary misorientations and geometries in real polycrystalline materials is questionable \cite{Zhang2017,Li2022}. Existing mobility measurements for recrystallization boundaries typically rely on averaging across multiple grains using the average driving force over the entire sample, thus concealing the variability in mobility among different boundaries \cite{Vandermeer1997,Huang1999,Taheri2004,Lens2005}. Recently, average mobilities have been quantified for individual recrystallization boundaries through \textit{ex situ} experiments that measure local boundary migration \cite{Zhang2018,Basu2016,Huang2012,Zhang2024}. However, as a recrystallizing grain grows into the deformed matrix, misorientation varies for individual dislocation cells along the boundary. The effect of this misorientation variation on boundary mobility and local boundary migration has not been thoroughly investigated.

The aim of this work is to develop a new method, termed Iterative Mobility Matching (IMM), for quantifying the local mobility variation along individual recrystallization boundaries. The method uses experimental data as input for a phase field simulation. This simulation is used to fit mobility values for each boundary segment of the recrystallizing grain encountered during the migration process, ensuring that the simulated migration aligns with the experimental boundary migration. This method is inspired by our previous works on determining materials parameters from \textit{in situ} experiments for the liquid diffusion coefficient \cite{Zhang2017a} and reduced grain boundary mobility during grain growth \cite{Zhang2020}. However, it must be emphasized that the present method differs fundamentally from these earlier approaches. It utilizes unique aspects of recrystallization, where a stored-energy driving force must be considered in addition to the curvature driving force. More critically, a new approach for projecting interfacial mobilities across the entire computational domain \cite{Zhang2023} is incorporated, preventing the need to consider each grain or subgrain as a separate order parameter. Furthermore, an efficient optimization algorithm is proposed without solving the computationally expensive sensitivity equations.

We first validate the IMM method using synthetic data and then apply it to time-series experimental data that investigated the migration process of a general high-angle recrystallization boundary during \textit{ex situ} annealing of pure aluminum \cite{Zhang2018}. This material was selected as a model material to establish and demonstrate the IMM approach. It offers a well-characterized and relatively simple system, free from complexities such as second-phase particles or solute effects, allowing us to focus on validating the methodology itself. Additionally, the microstructural and kinetic data were already available from prior experiments, which facilitated the integration of experimental and modeling efforts within the IMM framework. The mobility values of each pixel swept by the recrystallization boundary are then determined using the proposed method. The result shows that local mobility can vary by more than three orders of magnitude along a single recrystallization boundary. The mobility generally correlates well with misorientation, but the pattern is rather complex, indicating the effect of other factors. 

\begin{figure*}[t!]
  \centering
  \includegraphics[width=0.9\textwidth]{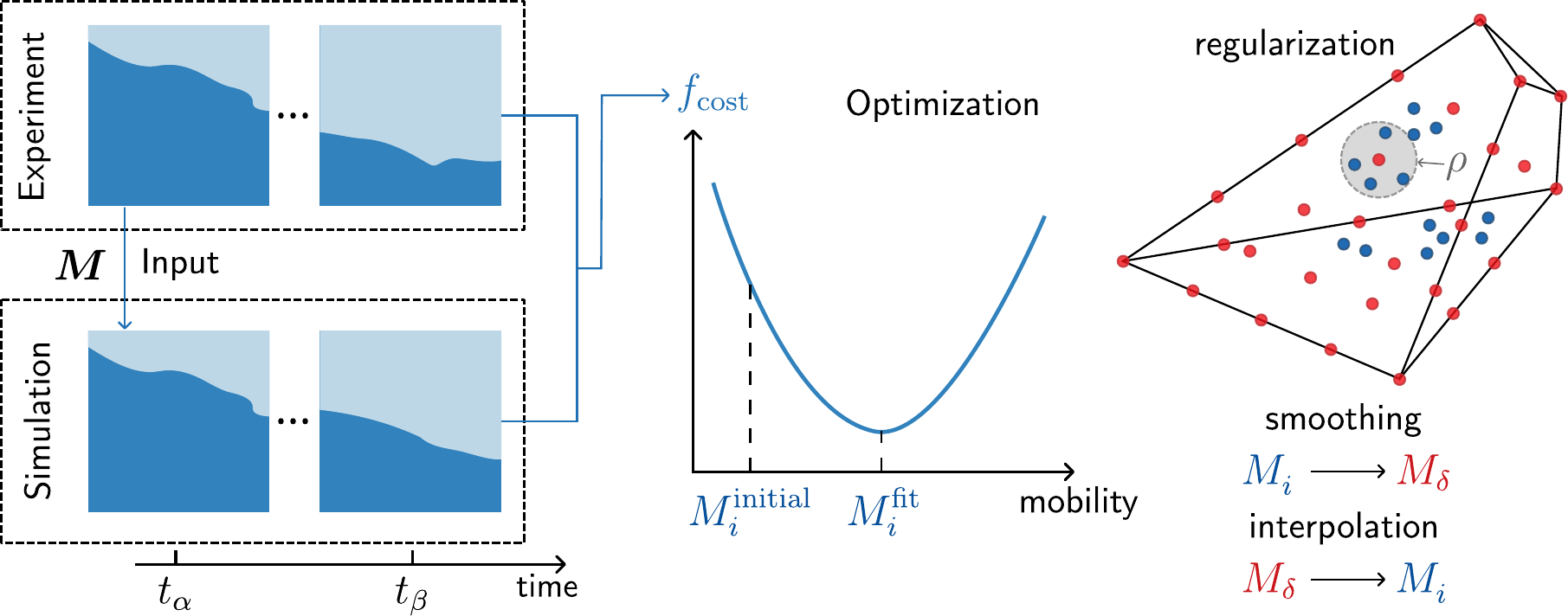}
  \caption{Schematic illustration of the Iterative Mobility Matching (IMM) method. The evolution of the recrystallized boundary is measured experimentally. Taking one experimental snapshot ($t_\alpha$) as input, the evolution of the recrystallized boundary is simulated with a guess of the mobility parameters $\bm{M}$ and compared with the experiment at $t_\beta$ to define a cost function $f_{\text{cost}}$, quantifying their dissimilarity. An optimization algorithm minimizes the cost function and determines the mobility $M_i^{\text{fit}}$ that best matches the simulation with the experiment. The fitted mobilities $M_i$ are regularized in the Rodrigues-Frank space using sampling points (red points). Data points (blue points; related to each pixel) within a radius $\rho$ to a sampling point in the misorientation space are used for regularization. The regularized mobilities $M_\delta$ are then interpolated to update the mobility guess at each pixel $M_i$. The whole process is repeated until convergence.}\label{fig:fitting:scheme}
\end{figure*}

\section{Methodology}\label{sec:method}
This section presents the IMM method, which consists of three main components: experimental data (boundary migration, stored energy and misorientation), a phase field model, and an optimization method. These components are illustrated in Fig.~\ref{fig:fitting:scheme}. While the experimental data will be detailed in subsequent sections, the latter two components are covered here, with a focus on the fundamental elements of the optimization method. These elements include the design variable, the objective function, the sensitivity analysis, and the optimization algorithm.

\subsection{Phase field model for recrystallization}\label{sec:phasefield}
This work uses a driven Allen-Cahn equation to simulate the recrystallization process. The boundary migration is modeled using the following phase field equation
\begin{equation}\label{eq:phasefield}
  \dpd{u}{t} = \mathcal{P}_u(M) \left(\frac{p'(u)}{6 \ell} F_{\text{s}} + \sigma\left(\lapl{u} - \frac{1}{2 \ell^2} g'(u) \right) \right),
\end{equation}
where $M(\vec{x})$ is the mobility, $u(\vec{x},t)$ is the phase field variable, which equals one in the deformed phase and zero in the recrystallized phase, with a smooth transition at the interface over a length scale $\ell$ referred to as the diffuse interface width, $g(u)=u^2(1-u)^2$ is a double-well potential, $p(u)=u^2(3-2u)$ is an interpolation function. $F_{\text{s}}(\vec{x})$ has the same meaning as in Eq.~\ref{eq:sharp}. Here, the mobility $M(\vec{x})$ is defined as the mobility if the recrystallization boundary is located at $\vec{x}$. Therefore, in Eq.~\ref{eq:phasefield}, we should only use the mobility where the boundary is currently located. To account for this, we introduce a projection operator $\mathcal{P}_{u(\vec{x},t)}(\cdot)$ to map the mobility at the current boundary (given by $u(\vec{x},t)=0.5$) to every point in the computational domain by making $\mathcal{P}_u(M)$ a constant perpendicular to the boundary. An example is given in Supplementary Fig.~S1. Note that this projection is needed for each time step. More details on the projection operator can be found in Ref.~\cite{Zhang2023}. Equation~\ref{eq:phasefield} can be seen as a diffuse interface version of the sharp interface model in Eq.~\ref{eq:sharp}. The derivative of the interpolation $p'(u)$ in Eq.~\ref{eq:phasefield} can be viewed as a regularized Dirac-delta function to localize the driving force close to the interface. The term in the bracket after $\sigma$ can be viewed as the diffuse interface version of the total curvature. By using Eq.~\ref{eq:phasefield} instead of Eq.~\ref{eq:sharp}, we prevent the need to track the interface explicitly. Moreover, it is unnecessary to calculate $v$ and $\kappa$ explicitly as they are embedded into the phase field equation. One key advantage of the IMM method is that the phase field model inherently captures the spatial variations in the thermodynamic driving forces, including stored energy and curvature, thereby enabling isolation of the dependence of mobility on misorientation.

\subsection{Optimization problem}\label{sec:optimizationproblem}
The goal of the IMM method is to find a set of mobilities, denoted as $\bm{M} = \{M_\delta|\delta=1,\cdots,n\}$, that minimizes the difference between the simulated boundary migration path $u(t_k;\bm{M})$ and the experimentally measured time series of boundary traces. These optimal mobilities are regarded as the physical materials parameters. The experimental microstructure is described by the signed distance functions $\phi_{\text{exp}}(t_k)$, where $t_k$ is the $k$-th experimental time step. The optimization problem is defined as
\begin{eqnarray}
    \text{find} & \bm{M} = \{M_1,M_2,\cdots,M_n\}\nonumber\\
    \text{minimize} & f_{\text{cost}}\left(u(t_k;\bm{M}),\phi_{\text{exp}}(t_k)\right)\label{eq:optimization}\\
    \text{such that} & G\left(u(t_k;\bm{M}), \bm{M}\right) = 0,\nonumber
\end{eqnarray}
where $f_{\text{cost}}$ is the objective function that measures the difference between the simulation and the experiment, which will be discussed later. The constraint function is
\begin{equation}
  \label{eq:optimization:constraint}
  G\left(u(t_k;\bm{M}), \bm{M}\right) = u(t_k) - u(\phi_{\text{exp}}(t_0)) - \intj[t_k]{t_{0}}{\dpd{u}{t}}{t},
\end{equation}
which enforces that $u$ satisfies the phase field equation (Eq.~\ref{eq:phasefield}).

\subsection{Design variables}\label{sec:designvariable}
After discretization of Eq.~\ref{eq:phasefield}, a mobility $M_i=M(\vec{x}_i)$ is assigned to each cell, where a cell corresponds to the computational grid. Here, $\vec{x}_{i}$ is the location of the \textit{i}-th cell. In this work, the computational grid size matches the experimental resolution, so the cell also represents a pixel/voxel in the 2D/3D experimental image. Though $M_i$ seems to be a natural choice of the design variable, it results in a huge design space that can easily lead to overfitting. Therefore, instead, we define the design variable $M_\delta$ at given sampling points in the misorientation space, as shown in Fig.~\ref{fig:fitting:scheme}. This approach helps to regularize the problem by using mobilities of fewer sampling points (red points in Fig.~\ref{fig:fitting:scheme}).\\
To interchange the mobilities defined at sampling points $M_\delta$ with those at each cell $M_i$ (referred to as data points), we introduce the following smoothing and interpolation processes
\begin{subequations}\label{eq:M}
\begin{equation}\label{eq:M:pixel2sample}
M_\delta = \frac{\sum_{\abs{\vec{R}_{i} - \vec{R}_\delta}<\rho} w_{i} M_i}{\sum_{\abs{\vec{R}_{i} - \vec{R}_\delta}<\rho} w_{i}},
\end{equation}
\begin{equation}\label{eq:M:sample2pixel}
M_i = \sum_{\delta=1}^n w_\delta (\vec{R}_{i}) M_\delta,
\end{equation}
\end{subequations}
where $\vec{R}_{\delta}$ is the misorientation of a sampling point, $\vec{R}_{i}=\vec{R}(\vec{x}_{i})$ is the misorientation of a data point determined experimentally, $w_\delta$ is the interpolation weighting factor, $w_{i}$ is the smoothing weighting factor, and $\rho$ is the radius of a sphere centered at the sampling point $\vec{R}_\delta$ in the misorientation space. The smoothing process includes all neighboring data points within the sphere of radius $\rho$, as shown in Fig.~\ref{fig:fitting:scheme}. In this work, linear interpolation is used in Eq.~\ref{eq:M:sample2pixel}, and $w_i$ is determined from the distance from this data point $\vec{R}_i$ to the sampling point $\vec{R}_\delta$: $w_i = \text{dist}(\vec{R}_\delta, \vec{R}_i)$.

\subsection{Objective function}\label{sec:objective}
The objective function quantifies the dissimilarity between the simulated microstructure represented by $u$ and the experimentally measured microstructure $\phi_{\text{exp}}$ at three levels:
\begin{subequations}\label{eq:fcost:list}
\begin{equation}\label{eq:fcost:local}
  f_{\text{cost}}^{\text{local},\alpha\beta}(\vec{x})=\abs{\phi_{\text{sim}}(u(\vec{x},t_\beta;\bm{M})) - \phi_{\text{exp}}(\vec{x},t_\beta)}^2,
\end{equation}
\begin{equation}\label{eq:fcost:ab}
  f_{\text{cost}}^{\alpha\beta} = \intj{\Omega_{\alpha\beta}}{f_{\text{cost}}^{\text{local},\alpha\beta}(\vec{x})}{V},
\end{equation}
\begin{equation}\label{eq:fcost}
  f_{\text{cost}} = \sum_{\alpha<\beta} f_{\text{cost}}^{\alpha\beta},
\end{equation}
\end{subequations}
where the local cost function $f_{\text{cost}}^{\text{local},\alpha\beta}$ is defined for each cell, the cost function $f_{\text{cost}}^{\alpha\beta}$ encompasses all cells for a given combination of experimental time steps $t_\alpha<t_\beta$, and the total cost function $f_{\text{cost}}$ comprises a given set of combinations of experimental time steps. The simulation starts from an initial condition given by $\phi_{\text{exp}}(t_\alpha)$, and the comparison with the experiment is performed at time $t_\beta$. $\Omega_{\alpha\beta}$ is the domain swept by the measured boundary between $t_\alpha$ and $t_\beta$, and the signed-distance function $\phi$ is related to the phase field $u$ by the hyperbolic tangent profile: $u(\phi) = \tfrac{1}{2}\left(1-\tanh \tfrac{\phi}{2\ell}\right)$.

\subsection{Sensitivity analysis}\label{sec:sensitivity}
Sensitivity is the gradient of a function in the design space. Sensitivity analysis is a key step in constructing an optimization algorithm because it provides information on how a function responds to small changes in each design variable. The sensitivity of the cost function derived using the adjoint method \cite{Bendsoe2003} is:
\begin{equation}
  \label{eq:optimization:sensitivity}
  \dod{f_{\text{cost}}^{\alpha\beta}}{\bm{M}} = - \dod{f_{\text{cost}}^{\alpha\beta}}{u(\vec{x},t_\beta)}  \left.\dpd{G(\vec{x})}{\bm{M}}\right\vert_{t=t_\beta},
\end{equation}
where the sensitivity of the constraint function $\partial{G}/\partial{\bm{M}}$ is determined from the following evolution equation,
\begin{subequations}\label{eq:optimization:sensitivity:evolution}
\begin{equation}\label{eq:optimization:sensitivity:evolution:equation}
  \dpd{}{t} \left(\dpd{G}{\bm{M}}\right) = - \dpd{M}{\bm{M}} \left(\frac{p'(u)}{6 \ell} F_{\text{s}} + \sigma\left(\lapl{u} - \frac{1}{2\ell^2} g'(u) \right) \right),
\end{equation}
\begin{equation}\label{eq:optimization:sensitivity:evolution:ic}
  \left.\dpd{G}{\bm{M}}\right\vert_{t=t_\alpha} = 0.
\end{equation}
\end{subequations}
The sensitivity equation (Eq.~\ref{eq:optimization:sensitivity:evolution}) includes $n$ equations, where $n$ is the number of sampling points, typically around 1000. Solving all the sensitivity evolution equations is computationally very expensive, if not prohibited. However, $\partial M/\partial \bm{M}$ is extremely sparse and localized: $\partial M(\vec{x})/\partial M_i = \delta(\vec{x}-\vec{x}_i)$, leading to the assumption that $M_i$ only affects the local cost function at $\vec{x}_i$. This assumption leads to a scaling property, which applies to each cell:
\begin{equation}
  \label{eq:scaling}
  f_{\text{cost}}^{\text{local},\alpha\beta}\left(t, k M_i\right) = f_{\text{cost}}^{\text{local},\alpha\beta}\left(kt, M_i\right),
\end{equation}
where $k$ is an arbitrary positive constant. This property is similar to the one used in our previous works \cite{Zhang2017a,Zhang2020}.

\subsection{Optimization algorithm}\label{sec:optimization}

The optimization algorithm is constructed around the scaling property (Eq.~\ref{eq:scaling}), which transfers the derivative with respect to the design variable (sensitivity) to a time derivative. As the scaling property is an approximation, an iterative optimization process is necessary to find the optimal mobilities. Using this idea, the optimization problem can be solved from the phase field equation, without resolving the time-consuming sensitivity evolution in Eq.~\ref{eq:optimization:sensitivity:evolution}. The optimization algorithm consists of the following steps
\begin{enumerate}
\item Initialize the optimization with an initial guess of the mobilities $M_\delta=M_i=M^{\text{initial}}$.
\item Solve the phase field equation (Eq.~\ref{eq:phasefield}) with the mobilities $M_i$.
\item Fit the mobility of each cell $M_i$ based on the scaling property (Eq.~\ref{eq:scaling}). The readers can refer to \cite{Zhang2017a,Zhang2018Phd,Zhang2020} for more details on the local fitting process.
\item Determine the mobilities at sampling points in the fundamental zone of the Rodrigues space $M_\delta$ from the mobilities of all neighboring cells $M_i$ by the smoothing given in Eq.~\ref{eq:M:pixel2sample}.
\item Update the mobility of each cell $M_i$ by interpolating the sampling points using Eq.~\ref{eq:M:sample2pixel}.
\item Go to step 2 and repeat this process until convergence.
\end{enumerate}

\section{Validation of the IMM method using synthetic data}\label{sec:synthetic}
The proposed IMM method is first validated using synthetic data generated by the phase field model, where ground truth mobility values are known.

\begin{figure*}[t!]
  \centering
  \includegraphics[width=0.82\textwidth]{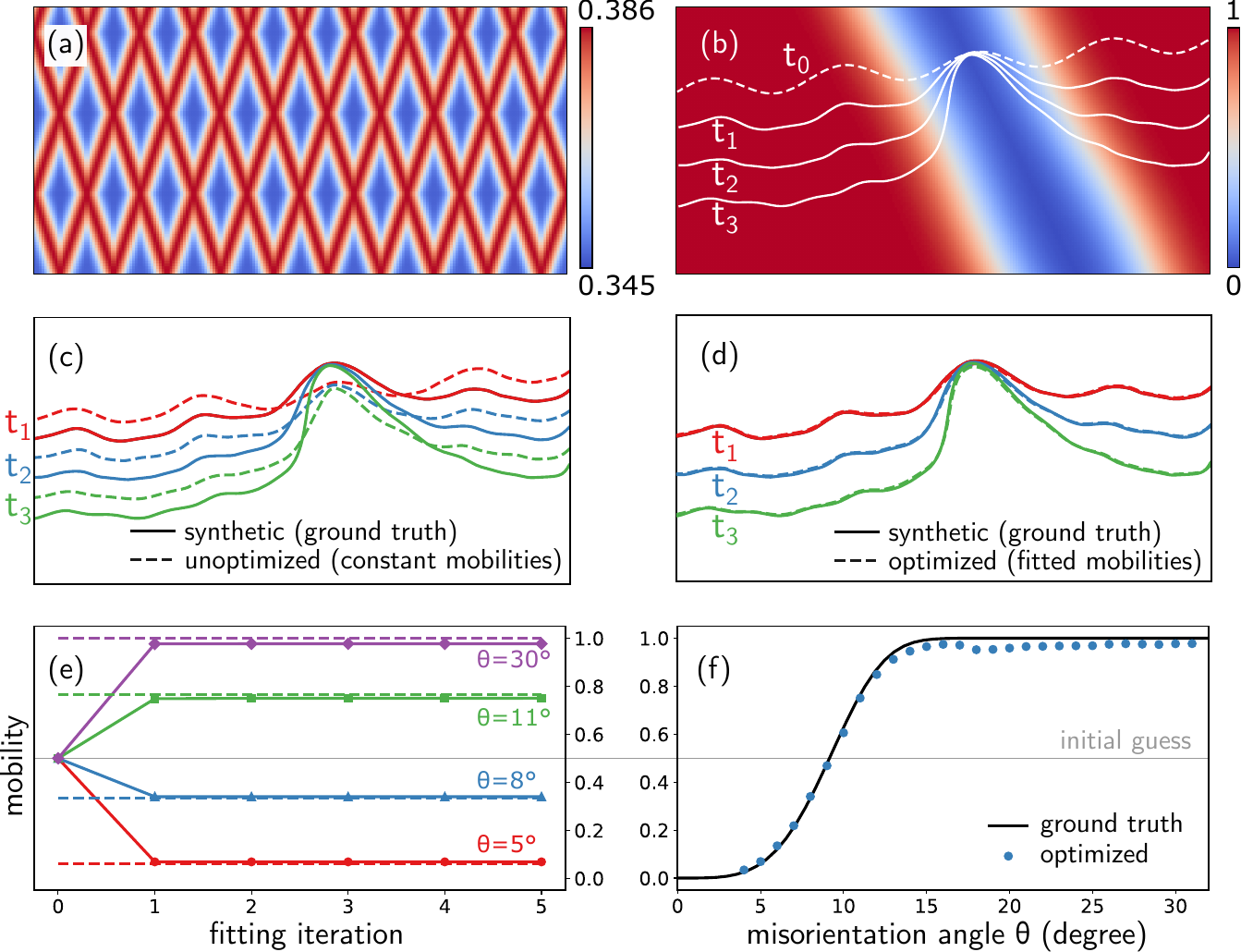}
  \caption{Validation of the Iterative Mobility Matching (IMM) method with synthetic data. (a) The synthetic stored energy. (b) The ground-truth mobility with overlaid boundary curves. (c) Simulated boundary evolution with the initial guess (unoptimized mobilities) by the dashed lines, along with the synthetic data (solid lines). (d) Comparison of the fitted boundary with the synthetic data. (e) Convergence curves of four sampling points with dashed lines representing the corresponding ground truth values. (f) Comparison of the fitted mobility (circular points) with the ground truth (solid line).}\label{fig:synthetic}
\end{figure*}
The synthetic stored energy is shown in Fig.~\ref{fig:synthetic}a. For simplicity, the misorientation is assumed to be defined by a fixed axis $\langle111\rangle$ and a misorientation angle $\theta$. We assume the mobility has the form \cite{Humphreys1997}
\begin{equation}
  \label{eq:synthetic:mobility}
  M = M_0 \left(1 - \exp \left[-B \left(\frac{\theta}{\theta_m}\right)^n\right]\right),
\end{equation}
where $B=5$, $n=4$, $\theta_m=15\si{\degree}$, and $M_0$ is the mobility of high-angle boundaries. The ground-truth mobility is depicted in Fig.~\ref{fig:synthetic}b. Starting with the initial boundary (dashed line in Fig.~\ref{fig:synthetic}b), we solve the phase field equation in Eq.~\ref{eq:phasefield} to generate the boundary trace with fixed time intervals (solid lines). To simplify the process for this test, the mobility and driving force are nondimensionalized by $M_0$ and $6\sigma/\ell$, respectively, equivalent to a constant scaling in time. The system consists of 100$\times$200 grids with a grid size of $1~\si{\um}$. We employ forward-Euler time integration with adaptive time step cutback to restrict the maximum change of the phase field variable in each time step to $0.01$. A linear extension boundary condition in \ref{sec:app:bc} is used.

For IMM, the phase field simulation is conducted with an initial microstructure at $t_\alpha$, and the fitting is performed at $t_{\alpha+1}$, where $\alpha=0,1,2$. For each combination $(t_\alpha,t_{\alpha+1})$, the mobility values for cells within the domain swept by the boundary from $t_\alpha$ to $t_{\alpha+1}$ are fitted and used for the regularization in Eq.~\ref{eq:M:pixel2sample}. We use a misorientation interval of 1 degree in the Rodrigues-Frank space to generate sampling points and a regularization radius $\rho=0.00467$ (approximately 0.5 degrees). With the initial guess of mobilities $M_\delta=0.5$, Fig.~\ref{fig:synthetic}c compares the simulated microstructures (dashed lines) and the synthetic `experiment' data (solid lines) at $t_1$ to $t_3$. After five optimization iterations, the IMM method improves the match between simulation and `experiment' as shown in Fig.~\ref{fig:synthetic}d. The convergence history of the fitted boundary traces is provided in Supplementary Video S1. Fast convergence is observed for four examples in Fig.~\ref{fig:synthetic}e (the dashed lines showing the ground truth), indicating the efficiency of the proposed optimization algorithm. Most mobilities are already close to their ground truth with just one iteration. Moreover, the IMM method is not sensitive to the initial guess, as shown in Supplementary Fig.~S2. The fitted mobilities as a function of the misorientation angle are shown in Fig.~\ref{fig:synthetic}f and compared with the ground truth given by Eq.~\ref{eq:synthetic:mobility}. A good match with the ground truth is observed, and the low- and high-mobility regions are correctly identified by the IMM method. This test validates the efficiency of the proposed IMM method in determining boundary mobilities.

\section{Extracting mobility values based on time-series experimental data}\label{sec:fitting}
We now apply the IMM method to the experimental data. To test the robustness of the proposed method, we will investigate the effect of various parameters, specifically the effect of filter size in processing the stored energy and the element size in discretizing the Rodrigues space.

\subsection{Recrystallization experiment}

\begin{figure*}[t!]
  \centering
  \includegraphics[width=0.82\textwidth]{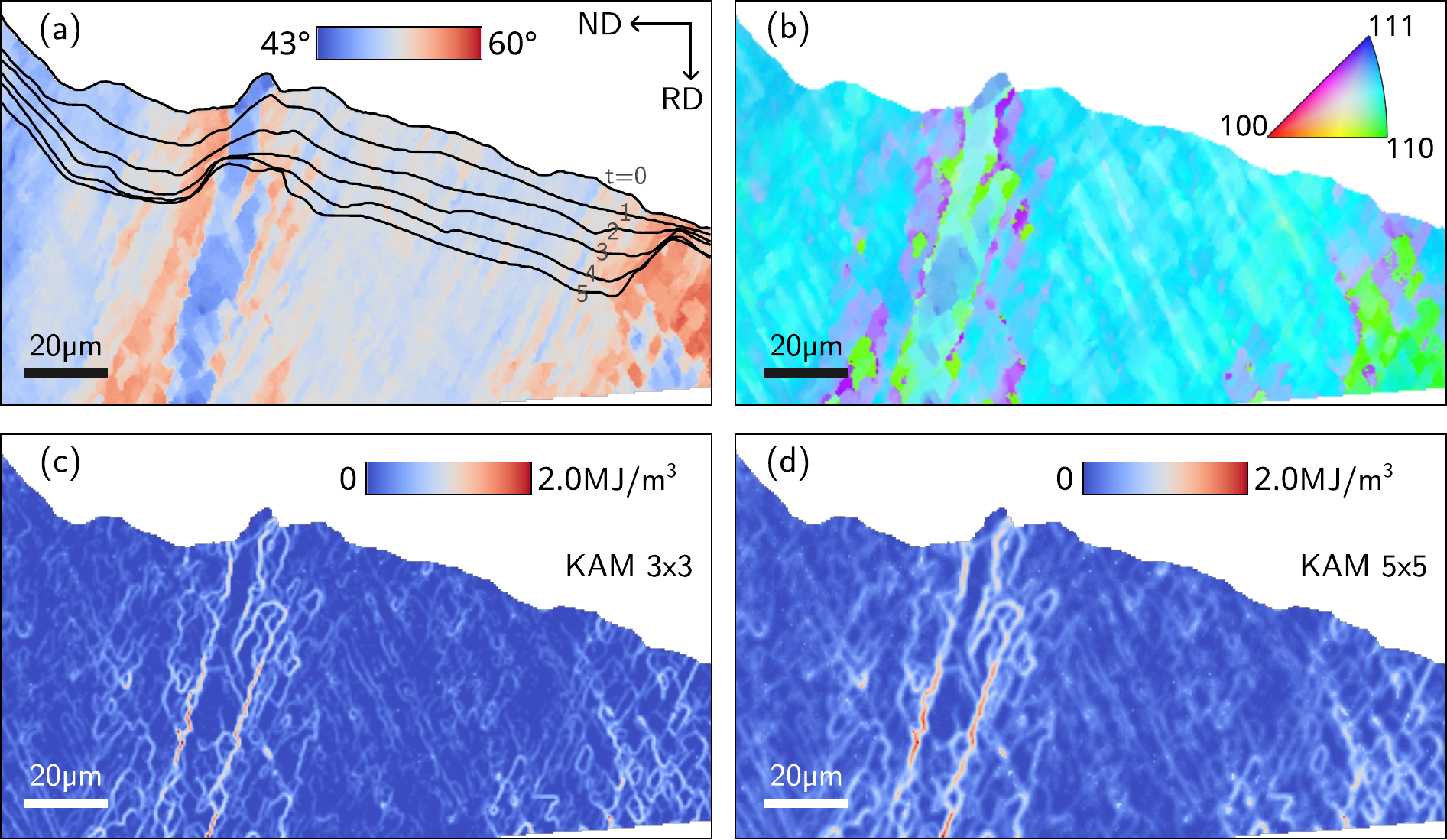}
  \caption{Recrystallization experiment. (a) Misorientation angle with boundary trace overlaid. (b) Misorientation axis. (c,d) Stored-energy map processed by KAM with (c) $3\times3$ and (d) $5\times5$ sized filters. RD: rolling direction, ND: normal direction.}\label{fig:experiment}
\end{figure*}

We use time-series experimental data following the migration of a recrystallization boundary during \textit{ex situ} annealing. A brief overview of the experiment is outlined below, with further details available in \cite{Zhang2018}. The experiment involved high-purity (99.996\%) aluminum with an initial grain size of several millimeters. The sample underwent a 50\% cold rolling deformation, followed by a recrystallization heat treatment at $250\si{\degreeCelsius}$ for 10 min in an air furnace. To identify a migrating recrystallization boundary, the sample was electropolished and annealed again at $250\si{\degreeCelsius}$ for 15 minutes. The microstructure around the selected boundary was characterized using electron channeling contrast (ECC) and electron backscatter diffraction (EBSD) techniques in a Zeiss Supra scanning electron microscope. However, due to the thermal and mechanical drift of the beam and sample, the EBSD map exhibited spatial distortion compared to the ECC image. This distortion was corrected using the method presented in  \cite{Zhang2014b} based on the ECC image of the corresponding area. The migration of the chosen recrystallization boundary was subsequently monitored using ECC in an \textit{ex situ} manner, with measurements performed in six steps following annealing at $250\si{\degreeCelsius}$ for 15 minutes. The initial microstructure and the traces of the migrating recrystallization boundary are shown in Fig.~\ref{fig:experiment}a. The misorientation angle and axis are determined from the EBSD map and shown in Figs.~\ref{fig:experiment}a and \ref{fig:experiment}b. The stored energy, $F_{\text{s}}$, input for the phase field simulation, was calculated using the EBSD map. For this calculation, the local stored-energy density for each pixel in the deformed matrix was determined using the method described in \cite{Zhang2017}, based on the kernel average misorientation (KAM) calculated with a subset of either $3\times3$ or $5\times5$ centered at the selected pixel. To reduce orientation noise, the EBSD map was first processed with a Kuwahara filter \cite{Humphreys2001,kuwahara1976digital}. All dislocation boundaries with misorientations $\geq{1\si{\degree}}$ were considered, with the boundary energy calculated based on the misorientation angle using the Read-Shockley equation \cite{Read1950}, where maximum energy of $\sigma_{\text{max}} = 0.324~\si{\joule\per\meter\squared}$ \cite{Murr1975} for boundaries with misorientations higher than 15\si{\degree} was used. The processed stored energies for the $3\times3$ and $5\times5$ filters are shown in Figs.~\ref{fig:experiment}c and \ref{fig:experiment}d, respectively. We will test the effect of different filter sizes on the fitting results.

\subsection{Phase field simulation setup}
The simulation domain consists of $190\times340$ grids with a grid size the same as the experimental resolution, $0.5~\si{\um}$. The adaptive time stepping method is the same as in the previous section. Since the recrystallization boundaries are high-angle (44.1 -- 57.6\si{\degree}) and far from special ($\Sigma$) boundaries, we assume a constant grain boundary energy of $\sigma=0.324~\si{\joule\per\meter\squared}$ \cite{Murr1975}. It's worth noting that our method is equally applicable to systems with anisotropic grain boundary energies. For example, anisotropic grain boundary energy can be reconstructed from experimental data \cite{Shen2019} and then incorporated into the phase field model \cite{Kobayashi1993,Wheeler1996}. A linear extension boundary condition (see \ref{sec:app:bc} for details) at all four outer boundaries is used to approximate the experiment.

\subsection{Regularization in the Rodrigues-Frank space}

\begin{table}[t!]
    \centering
    \caption{Regularization parameters for two sampling schemes.}\label{tb:regularization}
    \begin{tabular}{ccc}
    \hline
    Parameter name & Case 1 & Case 2 \\
    \hline
    Element size in Rodrigues space $e$ &  0.005& 0.01\\
    Regularization radius $\rho$ &  0.004 & 0.008 \\
    Number of sampling points & 2089 & 599 \\
    Average data points per sampling point &  11 & 40 \\
    \hline
    \end{tabular}
\end{table}

We consider each pixel swept by the recrystallization as an experimental data point. In total, there are 14580 data points. Using these data points directly as design variables can lead to overfitting. To prevent overfitting, we define the design variable on sampling points in the Rodrigues-Frank misorientation space. These sampling points are obtained by discretizing the Rodrigues space, with two average element sizes, $e=0.005$ and $e=0.01$, corresponding to misorientation angles of approximately $0.5\si{\degree}$ and $1\si{\degree}$, respectively. An example of the sampling points for $e=0.01$ is shown in Supplementary Fig.~S3a. Since the experiment data only covers a small portion of the Rodrigues space (blue points in Supplementary Fig.~S3b), we only consider sampling points with nearby data points (red points in Supplementary Fig.~S3b). The number of sampling points is much smaller than the data points, and a regularization radius $\rho$ is chosen to ensure a sufficient number of data points are related to each sampling point, as shown in Table~\ref{tb:regularization}. A regularization radius $\rho=0.008$ corresponds to a misorientation angle of about $0.9\si{\degree}$, close to the orientation resolution of the EBSD map after Kuwahara filtering, $1\si{\degree}$. Note that some data points close to the boundary of the $\rho$ sphere belong to more than one sampling point. To minimize the effect of double-counting, the distance between the sampling point $\vec{R}_\delta$ and the data point $\vec{R}_i$ is used to determine the weighting factor $w_i=a\, \text{dist}(\vec{R}_\delta,\vec{R}_i)$, with a prefactor $a$ reflecting the quality of the minimum found from the local fitting in the optimization algorithm (step 3). If no minimum is found, $a$ is set to 0, meaning this data point is not used in smoothing. If a minimum is found and its value is within $0.01-100$ of the previous mobility guess, $a$ is set to 1; otherwise, $a$ is set to 0.5. In addition, pixels within two grids from the outer boundary of the simulation domain have $a=0$ to minimize the effect due to boundary conditions. An example of the prefactor $a$ is shown in Supplementary Fig.~S4.

\begin{figure}[t!]
  \centering
  \includegraphics[width=.45\textwidth]{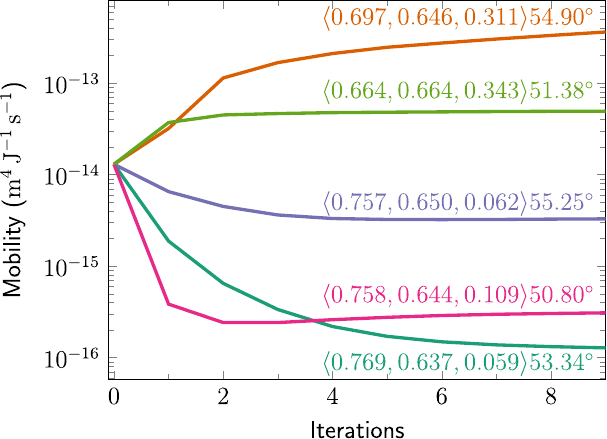}
  \caption{Convergence of the fitted mobility for selected sampling points.}\label{fig:convergence}
\end{figure}

\begin{figure*}[t!]
  \centering
  \includegraphics[width=0.8\textwidth]{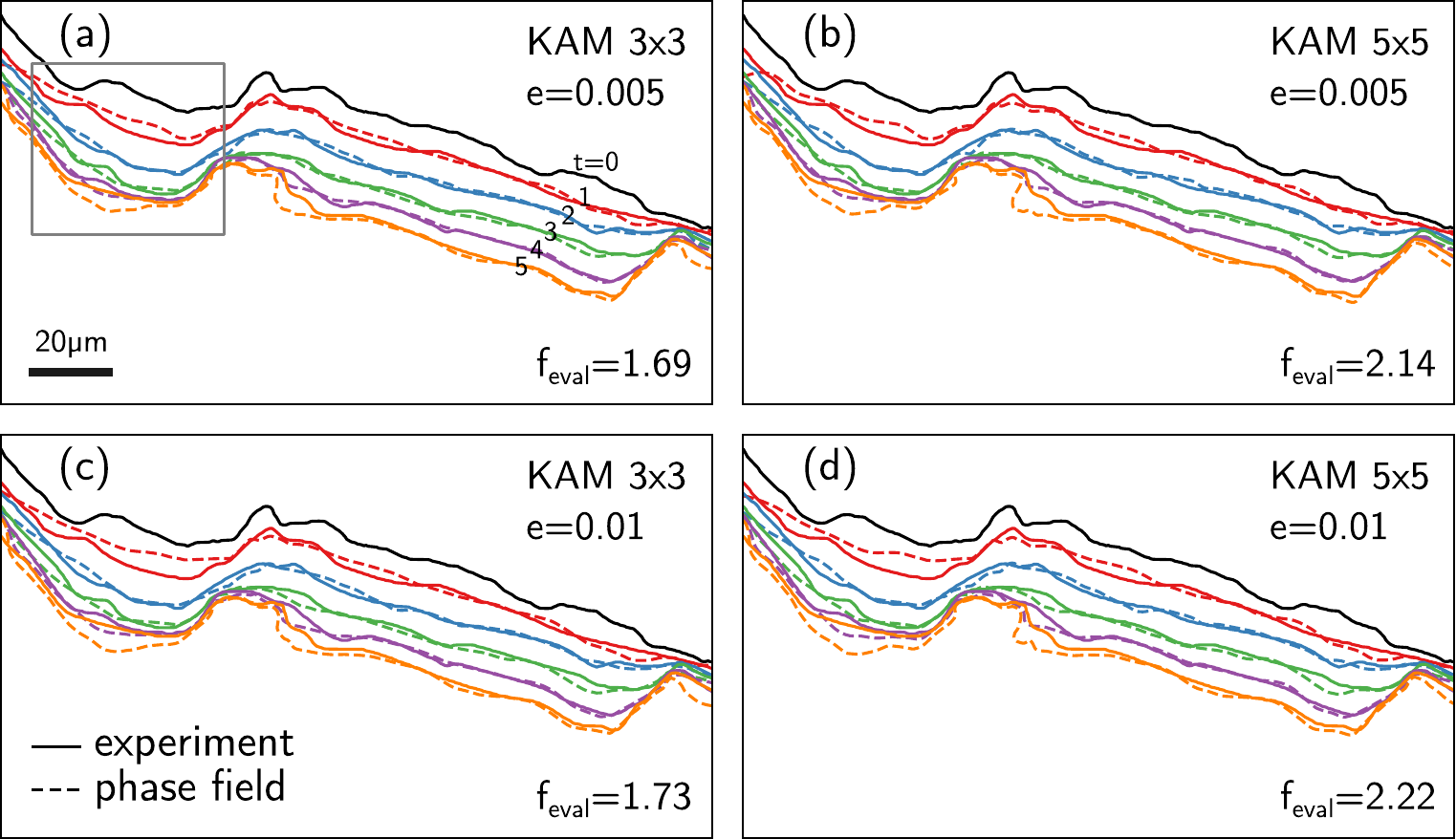}
  \caption{Comparison of fitted migration boundaries (dashed lines) with experimental measurements (solid lines). Each column corresponds to the stored energy processed by KAM with filters of $3\times3$ and $5\times5$. The first row employs a discretization of the Rodrigues space with an average element size of $e=0.005$, and the second row uses an average element size of $e=0.01$. The evaluation function $f_{\text{eval}}$ quantifies the quality of the fitting (lower is better). The unit of $f_{\text{eval}}$ is the grid size.} \label{fig:fitting}
\end{figure*}

\subsection{Implementation of IMM}
\begin{figure*}[t!]
  \centering
  \includegraphics[width=1.0\textwidth]{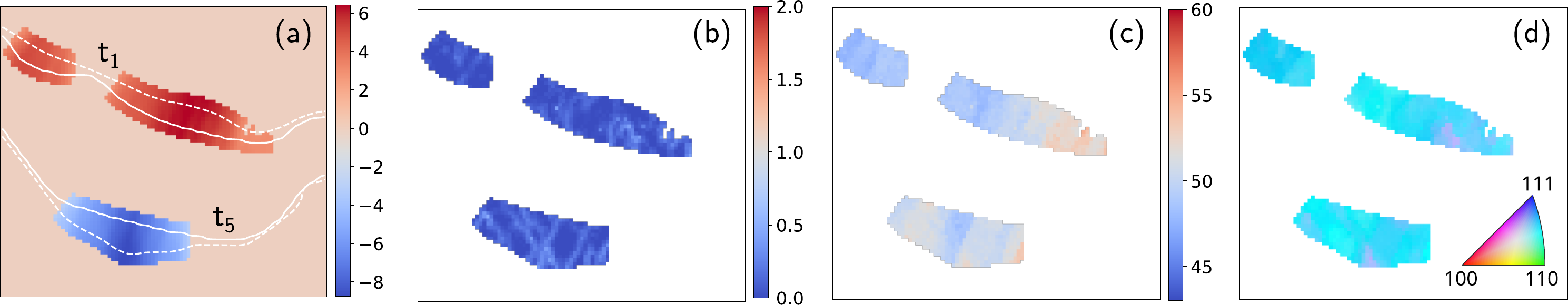}
  \caption{A closer view of the gray box region in Fig.~\ref{fig:fitting}a. (a) The distance between the simulated and experimental boundaries in units of the grid size. Only regions with a difference larger than five grids are shown. At $t_1$, the simulation (dashed line) under-predicts the experiment (solid line), while at $t_5$, the opposite is observed. (b) The stored energy in the corresponding region with unit \si{\mega\joule\per\meter\cubed}. (c) The misorientation angle in degrees. (d) The misorientation rotation axis.}\label{fig:mismatch}
\end{figure*}

We now apply the proposed IMM method to the experimental data in Fig.~\ref{fig:experiment}. The initial guess of the mobilities is $M^{\text{initial}} = 1.3\times10^{-14}~\si{\meter\tothe{4}\per\joule\per\s}$. A set of combinations of experimental time steps $\{(t_\alpha,t_\beta)|\alpha=1,\cdots,4;\beta=\alpha+1\}$ is used, where $t_\alpha$ is the starting time of the phase field simulation, and $t_\beta$ is when the fitting is performed. The adjacent time steps are chosen to minimize the effect of experimental uncertainty. The IMM took about two hours to run nine optimization iterations using an Intel Xeon Gold 6338 CPU (64 cores), with the majority of that time solving the phase field equation. The proposed algorithm shows fast convergence. Most mobility values converged after nine mobility matching iterations, with examples shown in Fig.~\ref{fig:convergence}. The optimization history of the fitted migration boundary is given in Supplementary Video S2. The convergence history of the mobilities in the Rodrigues space is shown in Supplementary Video S3. 

We also study the effect of filter size (used in processing $F_{\text{s}}$) and element size $e$ (discretization of the Rodrigues space). Two filter sizes, KAM $3\times3$ and KAM $5\times5$, and two element sizes, $e=0.005$ and $e=0.01$, are used, resulting in four cases. We compare the optimized boundary migration patterns with the experimentally measured ones in Fig.~\ref{fig:fitting} for the four cases and find a good match for most boundary segments. The match between the simulation and the experiment is characterized by the following evaluation function
\begin{equation}\label{eq:feval}
  f_{\text{eval}} = \frac
  {\displaystyle\sum_\beta \intj{\abs{\phi_{\text{exp}}(t_\beta)}<\varepsilon}{\abs{\phi_{\text{sim}}(t_\beta; \bm{M}_{\text{fit}}) - \phi_{\text{exp}}(t_\beta)}}{V}}
  {\displaystyle\sum_\beta  \intj{\abs{\phi_{\text{exp}}(t_\beta)}<\varepsilon}{}{V}},
\end{equation}
where the summation over $\beta$ accounts for all the experimental time steps used for the fitting, and $\varepsilon=3\Delta x$ specifies the integral domain as a narrow region around the experimental boundary. The quantity $f_{\text{eval}}$ has a physical meaning of the average distance between the simulated and measured boundaries and is shown in Fig.~\ref{fig:fitting} for all cases. The $f_{\text{eval}}$ values for each $t_\beta$ are listed in Supplementary Table~S1. Generally, small filter and element sizes provide a better fit. We notice some relatively large discrepancies for the boundary segments in the middle-left region, even for the best case (gray box in Fig.~\ref{fig:fitting}), with the largest mismatch at time steps 1 and 5, where the simulation under-predicts the experiment for $t_1$ and over-predicts for $t_5$. A zoom-in view of the region is shown in Fig.~\ref{fig:mismatch}.

Fig.~\ref{fig:mismatch}a shows the distance between the simulated and experimental microstructures $\phi_{\text{exp}} - \phi_{\text{sim}}(u)$, with a distance larger than three grids displayed. The inclinations of the boundary segments are similar between $t_1$ and $t_5$, and there is no significant difference in the stored energy, as given in Fig.~\ref{fig:mismatch}b. In addition, the misorientations for the boundary segments do not change much while the boundary is migrating along the rolling direction (RD), as shown in Figs.~\ref{fig:mismatch}c and \ref{fig:mismatch}d. This is further confirmed by comparing the misorientation in Rodrigues-Frank space between $t_1$ and $t_5$ shown in Supplementary Fig.~S5. The IMM method predicts the migration for the three intermediate time steps in the regions with similar store energies and misorientations; therefore, we suspect that under-prediction and over-prediction for the early and later annealing stages, respectively, might be due to other unknown physical driving forces and are not an issue of the IMM approach. One possibility is a different driving force due to local stress states at these two times. Overall, the IMM method shows that mobility is closely connected with misorientation, though the relationship is complex. In conclusion, the present IMM method works well for the experimental data. At the same time, it reveals a strong misorientation dependence of mobility, as the same mobility value can accurately describe boundary segments with the same misorientation but located in different regions along the boundary and observed at different annealing steps.

\subsection{Mobility results}

\begin{figure*}[t!]
  \centering
  \includegraphics[width=0.8\textwidth]{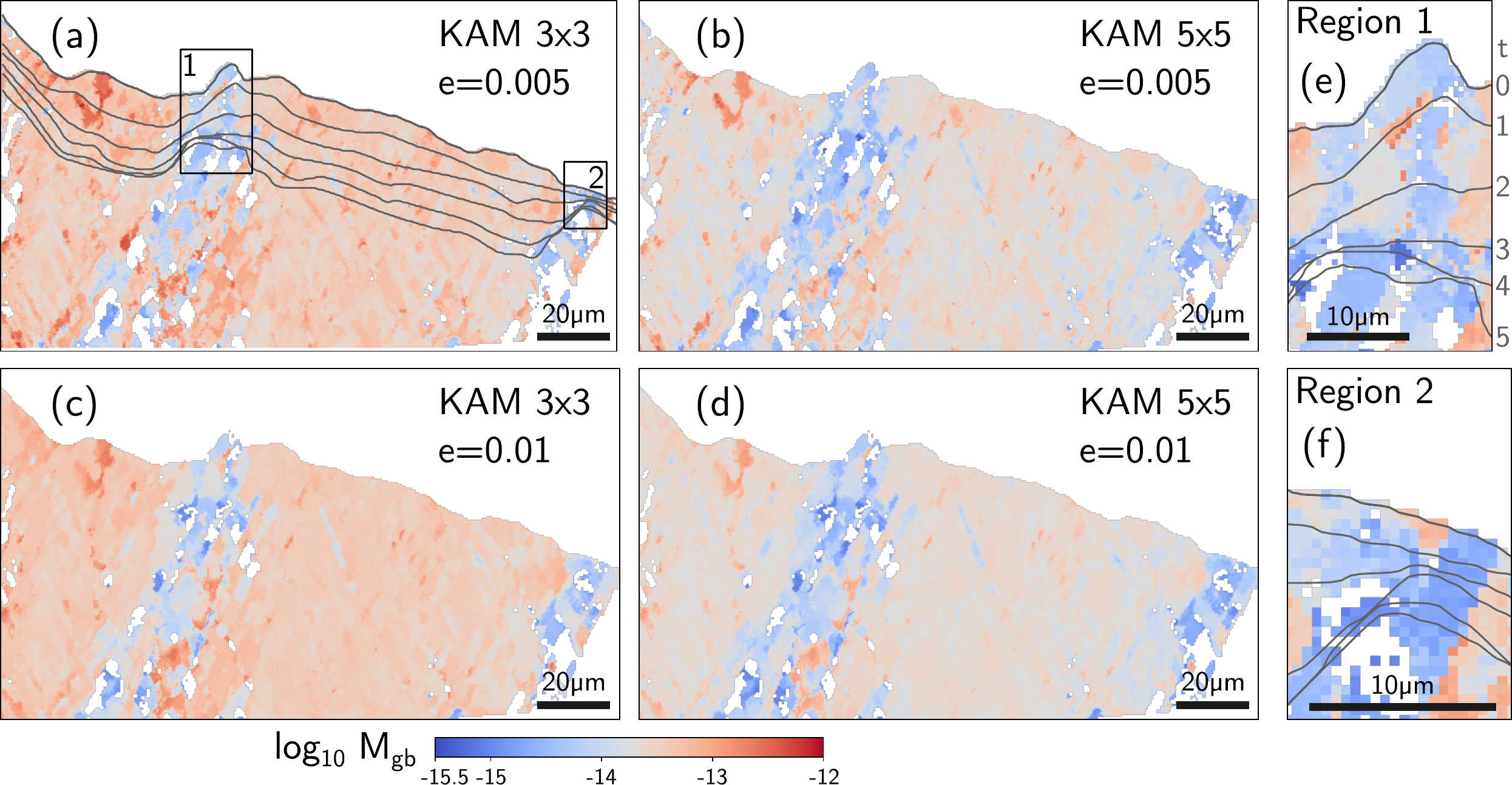}
  \caption{Comparison of fitted mobilities interpolated to each pixel. Each column corresponds to the stored energy processed by KAM with filters of $3\times3$ and $5\times5$. The first row uses a discretization of the Rodrigues space with an average element size of $e=0.005$, and the second row uses an average element size of $e=0.01$. (e,f) Zoomed-in view of the two low-mobility regions in (a). The empty pixels are not shown due to a low-quality fitting.} \label{fig:mobility:pixel}
\end{figure*}

\begin{figure}[t!]
  \centering
  \includegraphics[width=0.45\textwidth]{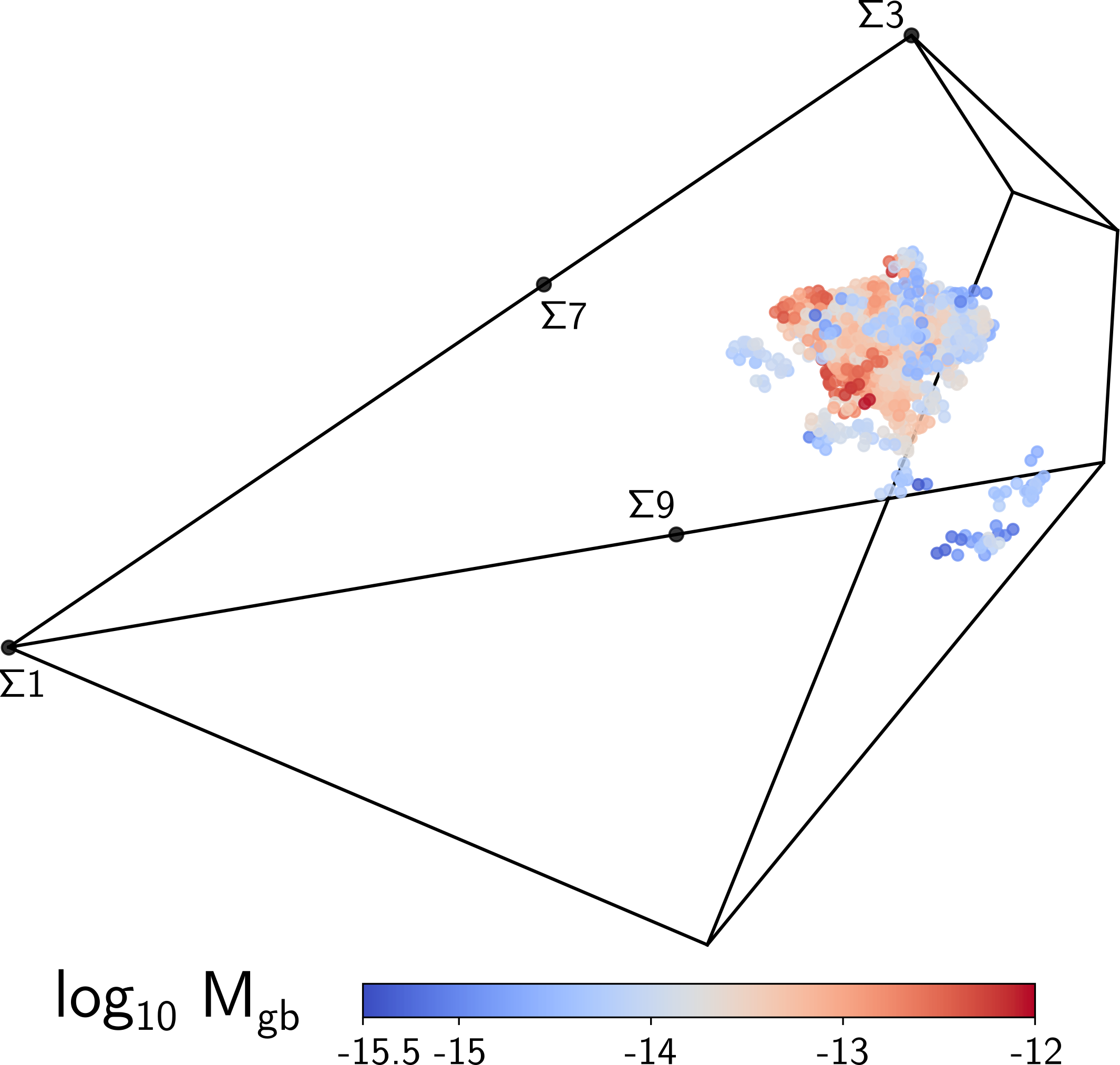}
  \caption{Fitted mobilities on sampling points in Rodrigues-Frank space for KAM $3\times3$ filter and an average element size of $e=0.005$.} \label{fig:mobility:rodrigues}
\end{figure}

Fig.~\ref{fig:mobility:pixel} shows the fitted mobility interpolated to each pixel of the microstructure. Points that are not converged or have high uncertainty (less than 3 data points inside the $\rho$-sphere) are not shown. Here, convergence is defined as the change in $\log(M_\delta)$ being less than 0.1. The fitted mobility in Rodrigues space is shown for KAM $3\times3$ and $e=0.005$ in Fig.~\ref{fig:mobility:rodrigues} (a 3D view is shown in Supplementary Video S4), with the remaining cases provided in Supplementary Fig.~S6. The IMM approach determined 1326 and 457 mobility values simultaneously for $e=0.005$ and $e=0.01$, respectively, showing a large mobility variation of about three orders of magnitude. The mobility variation patterns are not significantly affected by the filter and element sizes. As expected, the mobility variation decreases with increasing averaging effect in the Rodrigues space (larger $e$) or the real space (larger filter size). Although the absolute values differ slightly between the four cases, the IMM process identified the same high- and low-mobility regions. It is evident from the band appearance in Fig.~\ref{fig:mobility:pixel} that the mobility depends strongly on misorientation. However, the correlation between the two is complex, as revealed in Fig.~\ref{fig:mobility:rodrigues}. For better visualization, the highest and lowest mobility points are displayed in Supplementary Fig.~S7. The lowest values are mainly close to the $\langle110\rangle$ corner, while the highest values are relatively close to the $\langle111\rangle$ rotation axis (likely toward $\Sigma7$ boundaries). This trend agrees well with those obtained based on bicrystal boundary migration data \cite{Molodov1994}. The cluster of high-mobility boundaries with a misorientation angle of $52.7\si{\degree}$ is about $18\si{\degree}$ from the $\Sigma7$ boundaries, a well-accepted high-mobility boundary \cite{Huang1999}. No clear spatial correlation between the low mobility regions and the deviation angle to ideal $\Sigma7$ misorientation, see Supplementary Fig.~S8. This is likely because both angular and axis deviations contribute to the overall misorientation (see Figs.~\ref{fig:experiment}a and \ref{fig:experiment}b), and influence the boundary mobility differently.

\section{Discussion}\label{sec:discussion}

The results demonstrate that the proposed IMM method can reliably extract meaningful local mobility values for individual recrystallization boundary segments from 2D time-series experimental data, while also establishing a clear correspondence between boundary misorientation and mobility. For the first time, a substantial variation in local mobility, spanning approximately three orders of magnitude, has been observed. The implications of the observed local mobility variations on recrystallization boundary migration, along with further improvement of the IMM method, will be discussed in more detail.

\subsection{Effects of local mobility variations}
We observe a large local mobility variation, spanning about three orders of magnitude. This range is much larger than, but covers well the range of, the averaged mobility values determined by fitting average $v$ and $F$ data for boundary segments within subset areas of the deformed matrix in \cite{Zhang2018}: $6.0\times10^{-15} - 6.6\times10^{-14}~\si{\meter\tothe{4}\per\joule\per\s}$. By combining the experimental data (Fig.~\ref{fig:experiment}) and the measured mobilities (Fig.~\ref{fig:mobility:pixel}), we are now able to have a detailed analysis of the boundary migration behavior.

In Fig.~\ref{fig:mobility:pixel}a, the experimentally measured boundary traces are overlaid with the mobility result. The boxes highlight regions with low mobility. In these areas, the comparatively high stored energies in these regions (see Fig.~\ref{fig:experiment}) partially compensate for the low mobility of the boundary segments. Simultaneously, the low mobility leads to the formation of retrusions, which introduce additional curvature driving forces that help sustain the migration velocity in these regions, allowing these segments to maintain an average migration rate not far behind that of the high-mobility segments. The highest curvature driving forces at the retrusion tip are around $0.2~\mathrm{MJ/m^3}$, comparable in magnitude to the stored-energy driving forces. The greater the mobility difference, the larger the curvature of the retruded boundary must become to maintain coordinated migration over a long distance.

Moreover, notable differences are observed between the two regions: Region 2 exhibits stagnation during the first three time steps, whereas Region 1 does not. A closer inspection at Fig.~\ref{fig:mobility:pixel}a reveals a fundamental difference between these two regions: Region 2 mainly consists of a thicker band with low mobilities, while Region 1 contains thinner bands with a mix of high and low mobilities (see Figs.~\ref{fig:mobility:pixel}e and \ref{fig:mobility:pixel}f for a detailed view). The different migration behaviors suggest that the spatial variations in mobility may affect local boundary migration. Thinner bands result in mobility variations on a smaller scale, leading to small but sharp retrusions where the curvature driving force at their tips is sufficient for the low-mobility boundary segments to catch up. On the other hand, thicker low-mobility bands are more effective in pinning a larger fraction of the boundary, resulting in pronounced retrusions with low local curvature driving forces at their tips. It should be noticed that the local stored-energy variation also influences the recrystallization boundary migration, as revealed in Ref.~\cite{Moelans2013,Moelans2015}. The combined effects of complex mobility and stored-energy variations require further detailed study in the future.

\subsection{Further developments of the IMM method for recrystallization study}
The performance of the IMM approach relies on accurate, quantitative phase field models and high-quality experimental data. In this first application of the IMM method, \textit{ex situ} experimental data were used: EBSD maps from the initial state provided local stored-energy distributions and misorientation of the recrystallization boundary, while ECC imaging enabled tracking of boundary migration with comparatively high spatial resolution. High resolution EBSD \cite{britton2012} with improved angular and spatial resolution could be applied to reduce noise and uncertainty in the determined local stored energy and to improve the resolution of the misorientation space. High spatial and temporal resolution in ECC imaging would enable more precise tracking of boundary positions and curvature evolution, thereby reducing noise and uncertainty in the computed driving forces and migration velocities, and consequently in the derived mobility values. Other sources of uncertainty, such as errors in determining the driving force due to the lack of precise grain boundary energies and the neglect of elastic energy, should also be considered. These aspects may be improved by incorporating more accurate data from either advanced experiments \cite{Shen2019} or molecular dynamics (MD) simulations \cite{Olmsted2009,ratanaphan2015}. More importantly, EBSD maps collected at different stages of boundary migration should be carefully examined to assess whether recovery of the deformed matrix occurs significantly during annealing \cite{Yu2022}, and this should be incorporated into the phase field model. 

Secondly, the IMM method should be applied to a broader range of recrystallization boundaries, including general and special boundaries. One example is boundaries with misorientations close to the $\Sigma7$ orientation relationship \cite{Huang2012}. Experimentally, it has been shown that tilt boundaries with this misorientation tend to exhibit higher mobilities than their twist counterparts \cite{kohara1958}. Therefore, it is important to incorporate the boundary plane normal into the analysis. To achieve a comprehensive understanding of crystallographic effects, the IMM approach should be extended to \textit{in situ} 3D experiments that follow non-destructively the migration of the recrystallization front \cite{Zhang2017}. Ultimately, capturing the full 5D crystallographic character, encompassing both misorientation and boundary inclination, is essential to fully assess its influence on boundary migration behavior. With the recent upgrade to fourth-generation synchrotron source \cite{cloetens2025} and advancements in imaging techniques \cite{gursoy2024,yildirim2025}, it is now feasible to collect high-resolution (100 nm to submicron) 3D datasets over relatively large volumes, thus allowing sufficient experimental sampling across this 5D space. In the meantime, phase field model that can handle the full 5D crystallographic character is needed. In that case, the phase field simulations should be efficiently parallelized, using MPI or GPU-based computing, to manage the resulting large data volumes. 

Moreover, grain boundary mobility is strongly temperature-dependent \cite{Gottstein2009,homer2022} and influenced by solute drag \cite{Huang2012,kern2024} and second-phase particles \cite{Gottstein2009,anes2022}. It is therefore essential to extend the framework to study temperature dependence and more complex material systems in future work. This extension would require coupling with appropriate phase field models that incorporate additional physics, such as alloy systems with multiphases and multicomponents \cite{moelans2011quantitative} or solute drag effects \cite{steinbach2012phase}. To accurately inform such models, experimental data on composition fields and phase distributions, such as from 3D tomography or spectroscopy \cite{ge2022,mani2025}, would be necessary.

Furthermore, recent studies have revealed significant residual stresses present within recrystallizing grains in several material systems using synchrotron x-ray methods \cite{Zhang2022,Lindkvist2023,Zhang2024,lee2024}. These stresses result from the redistribution of the medium-/long- range residual stresses developed in the deformed matrix, and possibly from deformation incompatibility imposed by neighboring grains \cite{Lee2013,Kassner2013}. How these stresses influence local boundary migration needs to be clarified and incorporated into the model.

\section{Conclusions}\label{sec:conclusion}
This work presents an Iterative Mobility Matching (IMM) method for determining the local spatial variations of mobilities of recrystallized boundaries. Thousands of mobility values are determined simultaneously from a time-series recrystallization experiment and a complex correspondence between boundary misorientation and mobility has been established. The results indicate that local mobility can vary by three orders of magnitude for the same recrystallization boundary, affecting the boundary morphology.
This novel approach offers a unique opportunity to acquire a comprehensive set of material mobility data, contributing to a thorough understanding of the recrystallization process.

\section*{Acknowledgments}
YZ wishes to acknowledge the support of a research grant (VIL54495) from VILLUM FONDEN. This research was supported in part through the computational resources and staff contributions provided for the Quest high performance computing facility at Northwestern University which is jointly supported by the Office of the Provost, the Office for Research, and Northwestern University Information Technology.

\appendix
\section{Linear extension boundary condition}\label{sec:app:bc}
\setcounter{equation}{0}
For a discretization $u_{ij}$, the boundary condition at the $j=0$ boundary is
\begin{equation}
  \label{eq:apdx:bc}
u_{i0} = 2u_{i1}-u_{i2}.
\end{equation}
Note that we use a cell-centered finite difference method. The boundary condition for the remaining boundaries can be expressed similarly.

\bibliographystyle{elsarticle-num}

\end{document}